\documentclass[12pt]{iopart}
\usepackage{graphicx,latexsym}
\usepackage{dcolumn,url}
\usepackage{iopams}

\begin{document}

\title[Quantum magneto-electrodynamics of electrons embedded in a photon cavity]
      {Quantum magneto-electrodynamics of electrons embedded in a photon cavity}

\author{Olafur Jonasson$^{1}$, Chi-Shung Tang$^2$, Hsi-Sheng Goan$^{3,4}$
        Andrei Manolescu$^5$, and Vidar Gudmundsson$^1$}
\address{$^1$Science Institute, University of Iceland, Dunhaga 3,
        IS-107 Reykjavik, Iceland\\
        $^2$Department of Mechanical Engineering,
        National United University,
        1, Lienda, Miaoli 36003, Taiwan\\
        $^3$Department of Physics and Center for Theoretical Sciences,
        National Taiwan University,
        Taipei 10617, Taiwan\\
        $^4$Center for Quantum Science and Engineering,
        National Taiwan University,
        Taipei 10617, Taiwan\\
        $^5$Reykjavik University,
        School of Science and Engineering,
        Menntavegur 1, IS-101 Reykjavik, Iceland }
\ead{\mailto{olafur.jonasson@gmail.com}, \mailto{vidar@raunvis.hi.is},\mailto{cstang@nuu.edu.tw}}
%
%

\begin{abstract}
We investigate the coupling between a quantized
electromagnetic field in a cavity resonator and a Coulomb interacting 
electronic system in a nanostructure in an external magnetic field. 
Effects caused by the geometry of the electronic system and the polarization of the
electromagnetic field are explicitly taken into account. Our
numerical results demonstrate that the two-level system
approximation and the Jaynes-Cummings model remain valid in the weak
electron-photon coupling regime, while the quadratic 
vector potential in the diamagnetic part of the charge current leads to 
significant correction to the energy spectrum 
in the strong coupling regime. Furthermore, we find that a coupling to a strong cavity
photon mode polarizes the charge distribution of the system requiring a large
basis of single-electron eigenstates to be included in the model. 
\end{abstract}

\pacs{42.50.Pq, 73.21.-b, 78.20.Jq, 85.35.Ds}


\maketitle

%
%

\section{Introduction}

In the last decade there has been increasing interest in systems
capable of generating quantized fields containing a preset number of
photons. Manipulation of the state of scalable light-matter coupled
quantum systems is one of the key issues for their implementation
for optomechanical systems~\cite{Heinrich2010,Rabl2011} or quantum
information processing devices~\cite{Liu2001,Golubeva2011}. However,
searching clear evidence of light-matter coupling nonlinearity is
still a challenge. To this end, one has to reach a strong
light-matter coupling regime for optically driven systems in high
quality micro-cavities~\cite{Yoshie2004,Peter2005}, and demonstrate
its single-photon characteristics~\cite{Hennessy2007,Pres2007}.
Flexible experimental design of circuit quantum electrodynamics
offers great potential for practical device applications to explore
strong light-matter coupling at microwave
frequencies~\cite{Wallraff2004,Niemczyk2010,Hoffmann2010,Delbecq2011,Frey2011}.

To describe the interaction between matter and the photons of a
quantized electromagnetic field, the Jaynes-Cummings (JC) model
is often applied~\cite{Jaynes1963}. The JC model describes the 
interaction between a two-level system (TLS) and a single field mode.
It is a fundamental model in quantum optics and quantum information
science~\cite{Raimond2001}.
For a TLS with energy level spacing $\Delta E$, coupled with strength
${\cal E}_\mathrm{JC}$ to a resonator with photon energy
$\hbar\omega$, the JC model is valid when
both the detuning $\delta = |\hbar \omega - \Delta E|$ is
sufficiently small and the
light-matter coupling strength is much smaller than the photon energy
(${\cal E}_\mathrm{JC} \ll \hbar\omega$). The dynamics can then be
obtained by the JC model~\cite{Haroche2006} and the energy spectrum can be solved
exactly if the rotating wave approximation (RWA) is applied~\cite{Shore1993}.

Single modes of the electromagnetic field can be treated as the
population of a field oscillator with different Fock states 
(states with certain number of photons). It was discovered that a
three-level problem, called the coupled-channel cavity quantum
electrodynamics model~\cite{Wang1992}, can be exactly transformed to
a two-level one for arbitrary detuning~\cite{Wu1996}, in which the
eigenstates of energy and orbital angular momentum can be explicitly
expressed in terms of the Fock states.

More recently, utilization of the giant dipole moments of intersubband
transitions in quantum wells~\cite{Helm2000,Gabbay2011} has enabled
researchers to reach the ultrastrong light-matter coupling
regime~\cite{Ciuti2005,Devoret2007,Abdumalikov2008}.  In this
regime, the JC model is not applicable and the
coupling mechanism has to be explored beyond the
JC model~\cite{Zela1997,Sornborger2004,Irish2007}. Despite the above mentioned
experiments, a study of the coupling between electrons and cavity photons with a specified
nanostructure geometry in a perpendicular magnetic field is still
lacking.

In this work, we investigate the interplay of the dynamics of 
correlated electrons in a nanostructure to the quantum field of a
rectangular cavity resonator subject to an external magnetic
field.
By performing numerical computations we demonstrate how the electron-photon coupling
influences an electronic system embedded in a quantized photon field.
The TLS approximation and the JC model will be examined in both
the weak and the strong coupling regimes as well as the effects
of the diamagnetic part of the charge current in the 
electron-photon interaction term, which the JC model lacks.

%
\section{Model and Theory}\label{Sec:II}

The system under investigation is a two-dimensional electronic
nanostructure exposed to a quantized electromagnetic field of a
cavity resonator and a static (classical) external magnetic field at a 
low temperature. The electron-photon coupled
system can be described by the many-body Hamiltonian
\begin{equation} \label{HH}
H = H_{\rm e} + H_{\rm EM} + H_{\rm e-EM}\, ,
\end{equation}
where the first term describes the electronic system including the 
magnetic-field modified kinetic term $H_{\rm e}^0$ and a Coulomb
interaction term $H_{\rm Coul}$, namely

\begin{equation}\label{H_e}
H_{\rm e} = H_{\rm e}^0 + H_{\rm Coul} .
\end{equation}
The second term $H_{\rm EM}$ in Eq.\ (\ref{HH}) represents the
electromagnetic field in a cavity resonator and the third
term $H_{\rm e-EM}$ contains the coupling between the
electronic system and the quantized electromagnetic field.

The electronic nanostructure is assumed to be fabricated by split-gate configuration 
in the y-direction, forming a parabolic confinement with the characteristic 
frequency $\Omega_0$ on top of a semiconductor heterostructure. The ends of the 
nanostructure in the x-direction at $x = \pm L_x / 2$ are etched, forming a hard-wall confinement of length $L_x$. 
Thereby, a closed electronic narrow constriction is created in the $2$D electron gas. 
The external classical magnetic field is given by $\mathbf B=B\mathbf{\hat{z}}$ with a vector potential
$\mathbf A_{\rm ext}=(-By,0)$. Hence, $H_{\rm e}^0$ can be expressed in second quantization as
\begin{equation} \label{H_e0}
H_{\rm e}^0 = \int d{\bf r} \psi^\dag \left\{
\frac{\pi^2}{2m^*} + \frac{1}{2}m^*\Omega_0^2 y^2 \right\}
\psi \, ,
\end{equation}
where $\boldsymbol{\pi} = {\mathbf p}+ e{\mathbf A}_{\rm ext}/c$ is the mechanical momentum, 
$m^*$ is the effective electron mass, and
\begin{equation} \label{fields}
 \psi = \sum_{i}   \psi_i(\mathbf r) d_i, \  
 \psi^\dagger = \sum_{i}  \psi_i^*(\mathbf r) d^\dagger_i \\ 
\end{equation}
are fermionic field operators, with $d_i$ the annihilation- and
$d^\dagger_i$ the creation operator for an electron in the single-electron state
$|i\rangle$ corresponding to $\psi_i$.
Through out this paper we use Latin indices to 
label single-electron states (SESs)
as well as Hilbert state vectors and Greek subscripts to label many-electron Fock states (MESs).
Rationalized Gaussian units are used exclusively and $e$ denotes the positive elementary charge.
In Eq.\ (\ref{fields}), $\psi_i(\mathbf r)$ can be any complete orthonormal set of functions with the 
correct boundary conditions. However if we use the SESs of $H_{\rm e}^0$ as a basis, 
$H_{\rm e}^0$ is diagonal and simplifies to
\begin{equation} \label{H_e0-2nd}
 H_{\rm e}^0 = \sum_i E_i d_i^{\dag}d_i\, ,
\end{equation}
where $E_i$ is the energy of the SES $i$, associated to the eigenfunction $\psi_i(\mathbf r)$. 
The SESs are computed numerically  in a functional basis using a straightforward diagonalization method.

We can write the Coulomb interaction term in the second quantized form
\begin{equation} \label{H_Coul}
H_{\rm Coul} = \frac{1}{2} \sum_{ijrs} \langle ij|V_{\rm Coul} |rs
\rangle  d_i^{\dag} d_{j}^{\dag}d_{s}d_{r}\, ,
\end{equation}
where the Coulomb interaction potential can be written as
\begin{equation}
 V_{\rm Coul}(\mathbf r,\mathbf r') = 
 \frac{e^2/\kappa}{|\mathbf r-\mathbf r'|+\eta^2}
\end{equation}
with $\kappa$ denoting the relative dielectric constant of the
material and $\eta$ an infinitesimal convergence parameter.  The
Coulomb matrix elements in Eq.\ (\ref{H_Coul}) are thus expanded in the
basis of the SESs involving the integration with respect to the
observing location ${\mathbf r}$
\begin{equation}
 \langle ij|V_{\rm Coul} |rs
 \rangle = \int d{\mathbf r} \; \psi_i^*({\mathbf r}) {\cal I}_{jr}({\mathbf r})
 \psi_s({\mathbf r})
\end{equation}
and the integration with respect to the source location ${\mathbf r}^\prime$
\begin{equation}
 {\cal I}_{jr}({\mathbf r}) = \int d{\mathbf r}^\prime \psi_j^*({\mathbf
 r}^\prime) V_{\rm Coul}(\mathbf r,\mathbf r') \psi_r({\mathbf r}^\prime)\; .
\end{equation}
We use the SESs to construct many electron Fock states $|\mu\rangle$ which obey 
$H_{\rm e}^0 |\mu\rangle=E_\mu |\mu\rangle$ and employ an exact numerical diagonalization 
method to obtain, in the (noninteracting) basis $\{|\mu\rangle\}$, the
Coulomb interacting eigenvectors $|\mu)$ which satisfy 
$(H_{\rm e}^0+H_{\rm Coul}) |\mu)=\tilde E_\mu |\mu)$. We write the Coulomb 
interacting eigenvectors in terms of a unitary transformation~\cite{Yannouleas2007,Vidar2009}
\begin{equation}
 |\mu) = \sum_\nu \mathcal V_{\mu\nu} |\nu \rangle ,
\end{equation}
which is obtained in the diagonalization process. Then, in the new interacting MES basis 
$\left\{ |\mu) \right\}$, we can rewrite the full electron Hamiltonian  as
\begin{equation}
  \label{He_tot}
  H_{\rm e} = \sum_\mu |\mu) \tilde E_\mu (\mu| \ .
\end{equation}
The cavity electromagnetic field is described by the quantized vector potential 
$\textbf{A}$ in the radiation (Coulomb) gauge. The free field Hamiltonian is simply
\begin{equation}
 H_{\rm EM} = \hbar\omega a^{\dag}a \; ,
\end{equation}
where $\omega$ is the frequency of the resonant cavity mode,
$a^\dagger$ and $a$ being the creation and annihilation operators for photons.
The last term in Eq.\ (\ref{HH}) describing the interaction between the electrons and the
quantized electromagnetic field is given by
\begin{equation}  \label{H_e-EM}
 H_{\rm e-EM} = - \frac{1}{c}\int d{\mathbf r}\, {\mathbf j}_{\rm e} \cdot
 {\mathbf A} - \frac{e}{2m^*c^2} \int d{\mathbf r}\,  \rho_{\rm e} A^2 \,  ,
\end{equation}
in which the charge current density is defined by
\begin{equation}
 {\mathbf j}_{\rm e} = -\frac{e}{2m^*} \left\{ \psi^{\dag}(\boldsymbol{\pi}\psi)
 + ( \boldsymbol{\pi}^*\psi^{\dag})\psi \right\}\,  ,
\end{equation}
and the charge density $\rho_{\rm e} = -e\psi^{\dag}\psi$.  In this paper we will show 
that the $A^2$ term in Eq.\ (\ref{H_e-EM}) significantly affects
dynamical features in the strong electron-photon coupling regime.
The electronic nanostructure is placed in a rectangular cavity
forming an electromagnetic oscillator with hard-wall boundaries at
$-a_c/2 < x, y < a_c/2$ and $-d_c/2 < z < d_c/2$ with cavity volume
$V_c = a_c^2 d_c$. The proposed electromagnetic oscillator is a
single planar rectangular cavity with mutually locked dual
anti-phase outputs, in which the electronic nanostructure is oriented
in the $z = 0$ plane with the center at $(x,y)=(0,0)$. In the following, we will consider
only transverse electric (TE) modes ($E_z = 0$), where the electric
field $\textbf{E}$ is perpendicular to the direction of propagation.
The cavity supplies a monochromatic wave stabilized in the
$\textrm{TE}_{011}$ mode with longitudinally polarized electric
field along x-direction, or in the $\textrm{TE}_{101}$ mode
with transversely polarized electric field along y-direction.
In the Coulomb gauge, the vector potential of the electromagnetic field takes the form
\begin{equation}
 \label{Apot}
 \mathbf{A}(\mathbf r) =
 \left( {\begin{array}{*{20}c}
   \hat{\mathbf e}_x  \\
   \hat{\mathbf e} _y \\
 \end{array}} \right)
 {\cal A}
 \left[ {\begin{array}{*{20}c}
   \cos \left( \frac{\pi x}{a_c}\right)   \\
   \cos \left( \frac{\pi y}{a_c}\right)   \\
 \end{array}} \right]
 \cos \left( \frac{\pi z}{d_c} \right) \left( a + a^{\dag}\right)
\end{equation}
with the upper component denoting the $\textrm{TE}_{011}$ mode and the lower one 
representing the $\textrm{TE}_{101}$ mode. The Cartesian unit vectors are
$\hat{\bf e}_x$ and $\hat{\bf e}_y$.

We assume that the size of the cavity is much larger than that of the nanostructure, 
that is $L_x \ll a_c,d_c$. Utilizing this condition we can approximate the cosines in 
Eq.\ (\ref{Apot}) by unity and take $\mathbf A$ outside the integrals in Eq.\ (\ref{H_e-EM}) and obtain
\begin{eqnarray} \label{H_e-EM_2nd}
 H_{\rm e-EM} &=& {\cal E}_\mathrm{c} \sum_{i,j}  d_{i}^{\dag}
 d_{j}\; g_{ij} \left( a + a^{\dag} \right)\\
 &+& \frac{{\cal E}_\mathrm{c}^2}{\hbar\Omega_w}
 {\cal N}_{\rm e}
 \left[ \left( a^{\dag}a +  \frac{1}{2} \right) + \frac{1}{2}\left( a^{\dag}a^{\dag} + a a
 \right) \right]\nonumber ,
\end{eqnarray}
where $\Omega_w = \sqrt{\Omega_0^2+\omega_c^2}$ is the effective confinement frequency, 
$\omega_c = eB/(m^*c)$ the $2$D cyclotron frequency, 
${\cal E}_\mathrm{c} = e{\cal A} a_w \Omega_w  /c $ the characteristic energy for 
the coupling between electrons and cavity photons, $a_w = \sqrt{\hbar/(m^*\Omega_w)}$ 
the characteristic oscillator length scale, and $\mathcal N_e=\sum_i d_i^\dagger d_i$ 
the electron number operator. In Eq.\ (\ref{H_e-EM_2nd}) we use the dimensionless 
coupling $g_{ij}$ between the electrons and the cavity modes defined by
\begin{eqnarray}\label{g_ij}
 g_{ij} = \frac{a_w}{2\hbar} && \int d{\mathbf r} \left[ \psi_{i}^*({\mathbf
 r}) \left\{ (\mathbf{\hat e}\cdot
 \boldsymbol{\pi})\psi_{j}({\mathbf r}) \right\} \right. \nonumber \\
 && + \left. \left\{ (\mathbf{\hat e}\cdot \boldsymbol{\pi}) \psi_{i}({\mathbf r})
 \right\}^* \psi_{j}({\mathbf r})
 \right] ,
\end{eqnarray}
with $\mathbf{\hat e}\cdot\boldsymbol{\pi}=e_x\pi_x+e_y\pi_y$. 
The first and the second terms in Eq.\ (\ref{H_e-EM_2nd}) contribute, respectively, to the linear and
nonlinear optical excitation energy spectra, which will be explored later.

In our theoretical consideration, we will formally consider all the
higher-lying photonic modes, truncating the infinite matrix in order
to retain enough modes to reach sufficient convergence.  To obtain
convergent energy spectrum from the total Hamiltonian
Eq.\ (\ref{HH}), all the resonant and antiresonant terms in the
photon creation and annihilation operators will be taken into
account with arbitrary detuning. In the case of resonance, the
condition of the vacuum Rabi frequency to the cyclotron frequency
ratio larger than one implies that the higher photonic modes are
coupled to the transition, and the diamagnetic $A^2$ term of the
electron-photon coupling in Eq.\ (\ref{H_e-EM}) becomes
dominant~\cite{Hagenmuller2010}.

In the MES basis $\{|\mu ) \}$, we can rewrite the electron-photon interaction Hamiltonian as
\begin{eqnarray} \label{H_e-EM_MB}
H_{\rm e-EM} &=& {\cal E}_\mathrm{c} \sum\limits_{\mu\nu ij} {|\mu )
\langle\mu |{\cal V}^\dagger d_{i}^{\dagger} d_{j} {\cal V}|\nu\rangle (\nu |\; g_{ij}\left\{a+a^\dagger\right\} } \nonumber\\
      &+& \frac{{\cal E}_\mathrm{c}^2}{\hbar\Omega_w}
      \sum\limits_{\mu\nu j}
      |\mu )\langle\mu |{\cal V}^\dagger d_{j}^{\dagger}d_{j}{\cal V}|\nu\rangle (\nu | \nonumber\\
      && \quad\times \left\{\left( a^\dagger a+\frac{1}{2}\right) +
      \frac{1}{2}\left( aa+a^\dagger a^\dagger\right)\right\} .
\end{eqnarray}
The energy spectrum of the total Hamiltonian involving both the
electron-photon and electron-electron interactions has to be
obtained from the many-body space of the interacting electrons $\{ |\mu)
\}$ and the Fock-space of photons $\{ |M_{\rm ph}\rangle \}$, namely the uncoupled
electron-photon many-body states (MBSs) $|\breve{\mu}\rangle = |\mu)
\otimes |M_{\rm ph}\rangle$. Using the noninteracting electron-photon
MBSs, performing diagonalization of the total Hamiltonian
(\ref{HH}), the interacting electron-photon MBSs $|\breve{\mu})$ can be
expressed as

\begin{equation}
 \label{uniW}
 |\breve{\mu}) = \sum_{\nu}{\cal W}_{\mu\nu} |\breve{\nu} \rangle \, .
\end{equation}

It is important to note that, in arriving to Eq.\ (\ref{uniW}) we have performed 
truncations to a basis two times. The first one is when we only use the $N_{\rm SES}$ 
single electron states to construct our Fock-space basis $\{ |\mu \rangle \}$. 
The second is when we only use $N_{\rm MES}$ Coulomb interacting states $|\mu)$ 
and $N_{\rm EM}$ photon states to construct the joint photon-electron many-body basis 
$\{ |\mu)\otimes|M_{\rm ph}\rangle  \}$. A third truncation is likely to be needed if 
one needs to apply operations involving $\cal W$ many times, for example in time dependent 
calculations~\cite{EM-GME}. However in this paper, only static 
properties are calculated such as the energy spectrum and charge densities of the 
many-body eigenstates. Therefore a third truncation is not necessary.

The electron charge density operator $Q({\mathbf r})$ in the electron-photon coupled
system in the second quantized form is
\begin{equation}
Q({\mathbf r}) = -e \sum_{i,j} \psi_{i}^*(\mathbf r) \psi_{j}(\mathbf r)
d_{i}^{\dag} d_{j} .
\end{equation}
By taking trace of the operators in the coupled MBS space $\{
|\breve{\mu}) \}$, we obtain the correlated many-body charge
distribution $\langle Q({\mathbf r})\rangle$ for the electrons in the nanostructure 
interacting with the photon field
\begin{equation}
 \langle Q({\mathbf r})\rangle
 = \mathrm{Tr}\left\{ \rho(t) {\cal W}^{\dag} Q({\mathbf r}){\cal  W}\right\} .
      \label{Qxy}
\end{equation}
It should be mentioned that the total density operator $\rho(t)$
contains information of the interacting many-electron system and the
monochromatic photon modes as well.

Below we shall demonstrate our numerical results displaying tunable
dynamical interplay features between the interacting electrons and
the quantized photon field with either $x$ or the $y$
polarized electric components in an external magnetic field. We
mention in passing that in the construction of the JC model we assume only two
electron states to be active in the system. However, when we take a
realistic geometry into account, a sufficient number of electron
states will be used for guaranteeing numerical convergence.

\section{Results and Discussion}\label{Sec:III}

In order to explore the dynamical features of an electronic system
coupled to a single-mode quantum photon field, a simple TLS model is
often employed~\cite{Feranchuk1996,Li2009,Zhang2011}. The energy
spectrum of the TLS can be used to employ the JC model~\cite{Jaynes1963}.  Although the JC
model has been applied beyond the RWA, a rigorous analysis for the
validity of the JC model in a realistic system remains unexplored.
It is thus useful to consider an electronic nanostructure coupled
to a cavity photons taking into account realistic geometrical effects, by
performing numerical computation beyond the JC model for
comparison.

We assume that the electric nanostructure is fabricated by
GaAs-based semiconducting materials with electron effective mass
$m^*=0.067m_e$ and the background relative dielectric constant
$\kappa = 12.4$. The electronic nanostructure is modeled as an infinite
square potential well of length $L_x=300$~nm along the x-axis and transversely constricted by
split-gates with a parabolic confining strength $\hbar\Omega_0 = 1.0$~meV along
the y-axis.  A uniform and static magnetic field $B = 0.1$~T
is applied along the z-axis.  In order to obtain the energy
spectra of the closed electron-photon system, we have used
$N_{\rm MES}=200$ electron states and $N_{\rm EM}=20$ photon states.
The two electron MESs have been computed with $N_{\rm SES}=50$ 
and the single electron states with $N_{\rm SES}=N_{\rm MES}$. 
No calculations for three or more electrons are needed in this paper.

\begin{figure}[htbq]
      \centering
      \includegraphics[width=0.58\textwidth,angle=0,bb=0 40 330 657,clip]{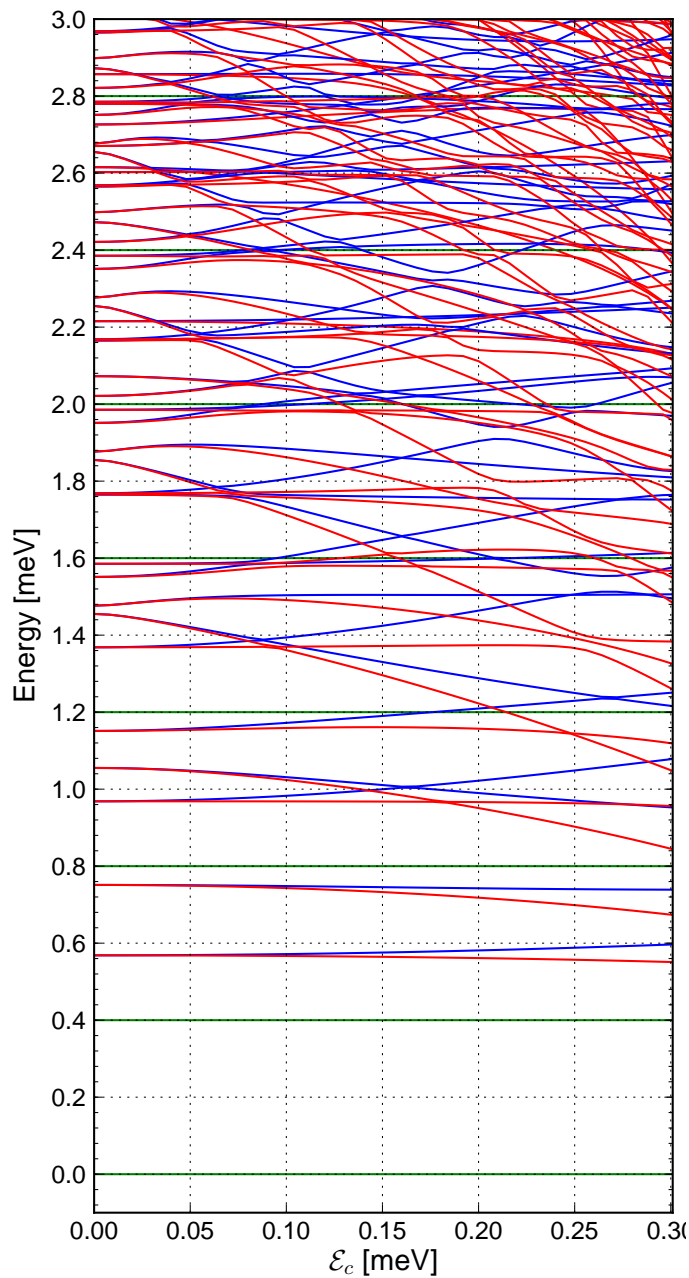}
      \caption{(Color online) The many-body energy spectra for the interacting-electron
               and the quantized-photon modes versus the electron-photon coupling strength
               ${\cal E}_\mathrm{c}$ for the case of $\textrm{TE}_{011}$ mode
               (the electric component with x-polarization).
               Shown are the pure photon states with no electron (green),
               the MBSs without the $A^2$ term (red), and the MBSs including the $A^2$ term
               (blue). The majority of the states contain a single electron. 
               Two-electron states
               are present for energy $> 2.0$~meV. Other parameters are
               $B=0.1$~T, $\hbar\Omega_0=1.0$~meV, and $\hbar\omega = 0.4$~meV.
               }
      \label{E-x}
\end{figure}

\begin{figure}[htbq]
      \centering
      \includegraphics[width=0.58\textwidth,angle=0,bb=0 40 330 657,clip]{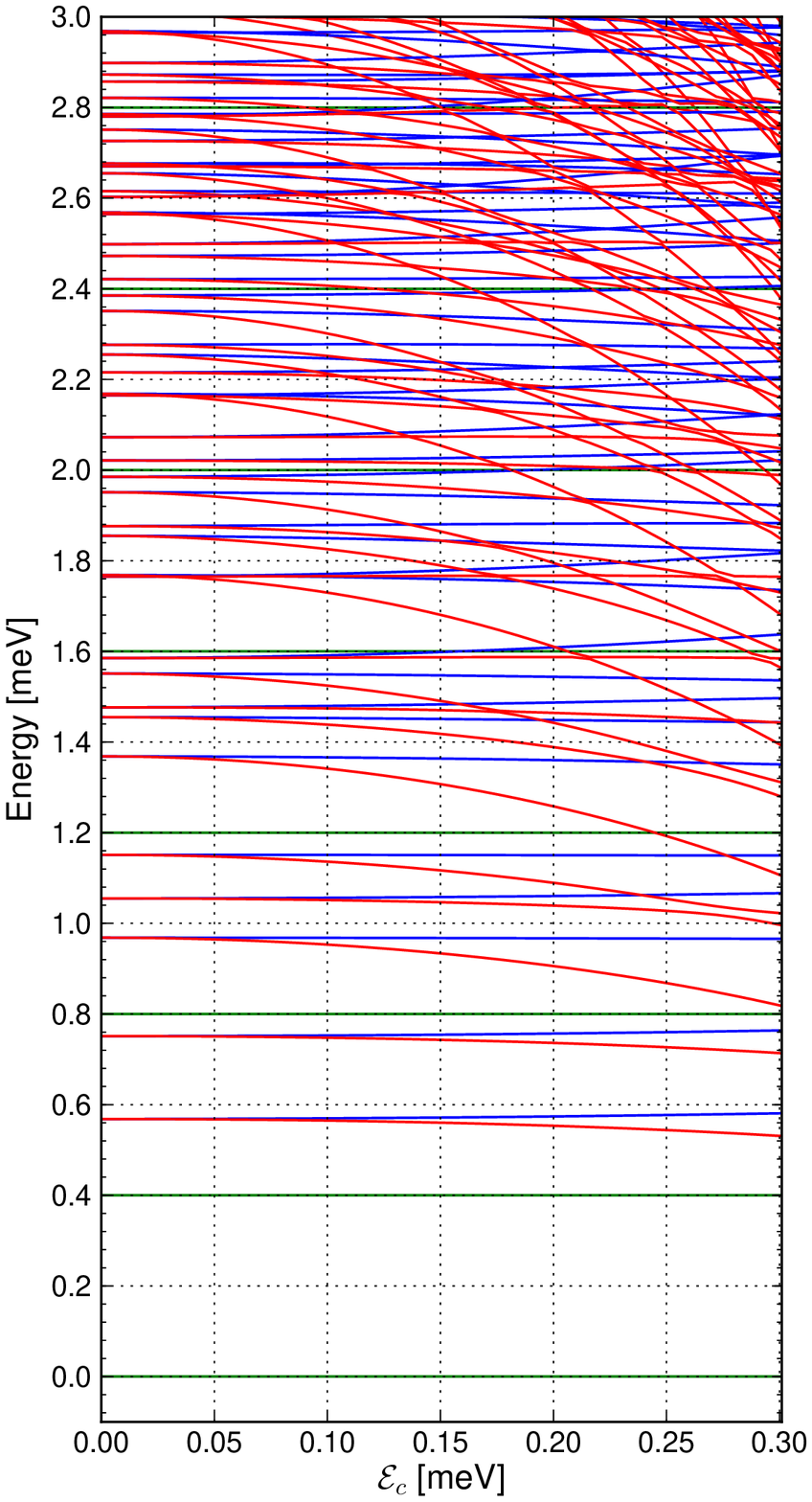}
      \caption{(Color online) The many-body energy spectra for the interacting-electron
               and the quantized-photon modes versus the electron-photon coupling strength
               ${\cal E}_\mathrm{c}$ for the case of $\textrm{TE}_{101}$ mode
               (the electric component with y-polarization).
               Shown are pure photon states with no electron (green),
               the MBSs without the $A^2$ term (red), and the MBSs including the $A^2$ term
               (blue). The majority of the states contain a single electron. 
               Two-electron states
               are present for energy $> 2.0$~meV. Other parameters are
               $B=0.1$~T, $\hbar\Omega_0=1.0$~meV, and $\hbar\omega = 0.4$~meV.
               }
      \label{E-y}
\end{figure}

The MBS energy spectra of the electron-photon states are shown in
Figs.\ \ref{E-x} and \ref{E-y} for the case of $x$ and
$y$-polarizations respectively, corresponding to the
$\textrm{TE}_{011}$ and the $\textrm{TE}_{101}$ modes. 
The single photon energy is $\hbar\omega = 0.4$~meV, hence the
$M_\mathrm{ph}$ photon states with no electrons
have the energy $E_\mathrm{ph} = M_\mathrm{ph} \hbar\omega$.
The horizontal axis ${\cal E}_\mathrm{c}$ denotes the strength of
coupling between the electrons and the photons. In the absence of
the electron-photon coupling ${\cal E}_\mathrm{c} =0$, both the
cases of $x$- and $y$-polarizations manifest the same
energy spectra.

In the weak electron-photon coupling regime ${\cal E}_\mathrm{c}
\leq 0.1 \hbar\omega$, the energy spectra of a linear approximation
neglecting the $A^2$ term (red) are only slightly shifted from the
full numerical results including the $A^2$ term (blue), and the pure
photon states (green) retain the same energy. However, the linear
approximation neglecting the $A^2$ term becomes inaccurate when the
coupling strength  ${\cal E}_\mathrm{c}$ is comparable to the
driving photon energy $\hbar\omega$ and the characteristic energy
level spacing of the electronic system.

Our full numerical results clearly show that the lowest two many-body 
energy states at around $E$ = $0.56$ and $0.75$~meV are not
sensitive to the polarization of the quantized electromagnetic
field. However, the higher MBSs may be sensitive to the polarization
of the photon field.  Moreover, comparing Figs.\ \ref{E-x} and
\ref{E-y}, we see that the energy dispersion is much more sensitive
to the $x$-polarized photon field than that of
$y$-polarization. This is because the single photon
energy $\hbar\omega$ is comparable to the characteristic energy of the
lowest states in the nanostructure for electron motion in the $x$-direction. 
In other words, it is caused by the anisotropy of the selected geometry. 
In addition, for the case of $x$-polarization the two states at around MBS
energy $E_{\rm MBS} \approx 1.0$~meV cross at ${\cal
E}_{\mathrm{c}}\approx 0.16$~meV, but not in the case of
$y$-polarization.

\begin{figure}[htbq]
      \centering
      \includegraphics[width=0.70\textwidth,angle=0]{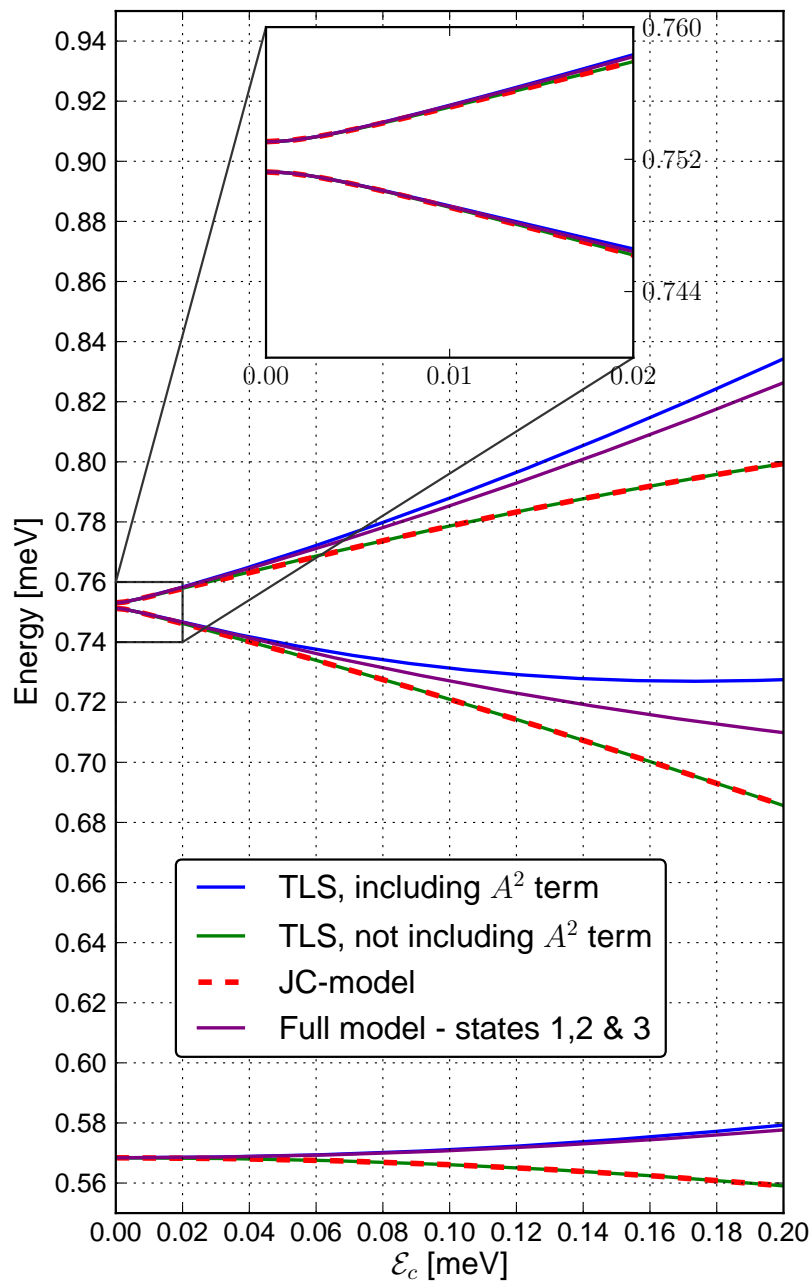}
      \caption{(Color online) Comparison of the many-body energy spectra
      versus the coupling strength $\mathcal{E}_\mathrm{c}$
      for the case of $\textrm{TE}_{011}$ mode ($x$-polarization).
      The energy spectra's are obtained by the TLS model with (blue) and without (green) the
      $A^2$ term and the JC model (red-dashed). The TLS model results are compared 
      with the full numerical calculation
      results for the lowest active levels $1$, $2$, and $3$ (purple).
      Other parameters are $B=0.1$ T, $\hbar\Omega_0=1.0$~meV, $|g_{12}|=0.701$,
      and $\hbar\omega = 0.185$~meV. The inset shows the validity of the JC model
      in the weak coupling limit.}
      \label{JC-xpol}
\end{figure}

We will now analyze the validity of the TLS approximation, as well
as the even simpler JC model. The JC model is
built from a TLS relying upon the assumptions of near resonance and
weak coupling between the two systems that is described by the following
Hamiltonian in the second quantized form
\begin{equation} \label{H_JC}
 H_\textrm{JC} = \frac{1}{2}\Delta E_{ij} \sigma_z + \hbar\omega a^{\dag}
 a + \mathcal{E}_\textrm{JC}\left( \sigma_{+} + \sigma_{-}\right)
 \left( a + a^{\dag} \right),
\end{equation}
where $\Delta E_{ij} = E_j - E_i$ denotes the energy difference between the electron states 
$|i\rangle$ and $|j\rangle$ which have been chosen as the relevant states for the 
TLS approximation. The ladder operators appropriate for a two-level approximation
$\sigma_{\pm}$ are defined by $\sigma_{\pm}
= \frac{1}{2}(\sigma_x \pm i \sigma_y)$, where $\sigma_{x,y,z}$ are the Pauli matrices. 
Note that the energies of states $|i\rangle$ and $|j\rangle$ have been shifted to make 
them symmetric around the zero energy.

The counter-rotating terms $\sigma_+ a^\dag$ and $\sigma_- a $ in Eq.\ (\ref{H_JC}) are usually 
omitted by taking the RWA to get an exactly solvable model. However, for our comparison 
we will keep the counter rotating terms and solve Eq.\ (\ref{H_JC}) numerically using the Fock space 
basis $\{ |k\rangle \otimes |M_{\rm ph}\rangle \}$, where $k\in\{i,j\}$. 
Comparison of the JC model with and without the counter rotating terms has been 
investigated~\cite{Feranchuk1996,Li2009,Zhang2011}, however it should be 
reexamined and compared with a system where realistic effects are included, such as those 
stemming from the non-trivial geometry of the nanostructure and an external magnetic-field.

For a TLS with one electron, zero magnetic field and ignoring the $\mathcal E_c^2$ 
term in Eq.\ (\ref{H_e-EM_MB}), the total Hamiltonian in Eq.\ (\ref{HH}) reduces to a JC-like Hamiltonian  
with a coupling strength  $\mathcal{E}_\mathrm{JC}$ associated with $\mathcal{E}_\mathrm{c}$ according to
\begin{equation} \label{E_JC}
 \mathcal{E}_\textrm{JC} = |g_{ij}| \mathcal{E}_\mathrm{c},
\end{equation}
in which the dimensionless coupling constant $g_{ij}$ can be calculated using Eq.\ (\ref{g_ij}). 
For a non-zero magnetic field Eq.\ (\ref{E_JC}) is not exact, however as will be shown later, 
a low magnetic field $B=0.1$~T has minimal effects.

Below we will assume that the detuning $\delta$ is one percent of   
the energy spacing of the two active states $\Delta E_{ij}$ giving the single-photon 
energy $\hbar\omega = 1.01 \Delta E_{ij}$. To label energy levels the notation $E_k^M$ 
is used. It refers to the energy of the state $| k \rangle \otimes |M_{\rm ph}\rangle$ 
for $\mathcal E_c=0$.

Figure \ref{JC-xpol} shows the $x$-polarization many-body energy spectra as a 
function of the electron-photon coupling strength for the different models. 
We consider the lowest states $|1\rangle$ and $|2\rangle$ as the relevant states 
for the TLS. In the zero coupling limit $\mathcal{E}_\mathrm{c} = 0$, the ground 
state energy $E_1=E_1^0 \simeq 0.568$~meV and the energy of the first excited state is 
$E_2=E_2^0 \simeq 0.751$~meV so that the energy level spacing is $\Delta E_{12} = 0.183$~meV. 
The detuning is small compared to the typical energy difference of the electron 
states so the third state is associated with one-photon absorption 
from state $|1\rangle$ with energy $E_3 = E_1^1 = E_1^0 + \hbar\omega = 0.753$~meV. 
As expected, the JC results almost coincide with our TLS results not including the $A^2$ term. 
The difference (not visible in Fig.\ \ref{JC-xpol}) between the two curves is due to effects 
of the external magnetic field. When the $A^2$ term is included, the energy spectrum manifests 
a blue-shift, and the energy-level correction $\delta E$ may be larger than $0.02$~meV in the 
strong coupling regime, or $\delta E / \Delta E > 10\% $. A weaker red shift correction is 
observed when the higher MBSs are involved in the electron-photon coupling.

In the weak coupling regime $\mathcal{E}_\mathrm{c} < 0.1\hbar\omega
\simeq 0.02$~meV, the JC model is approximately valid.  When the
coupling strength is increased to $\mathcal{E}_\mathrm{c} \simeq
\hbar \omega \approx 0.2$~meV, the ground state energy calculated by the 
TLS model is still valid. However,
the energy of the excited states becomes inaccurate, 
indicating that the simplified TLS model is no longer a good
approximation in the strong coupling regime even though the
diamagnetic vector potential $A^2$ is included.  When the coupling
strength $\mathcal{E}_\mathrm{c}$ is increased, both the JC model
and the TLS without the $A^2$ term predict a decreasing ground state,
however by including the $A^2$ term within the TLS model the energy increases,
in better agreement with our full numerical calculation.

\begin{figure}[htbq]
      \centering
      \includegraphics[width=0.68\textwidth,angle=0]{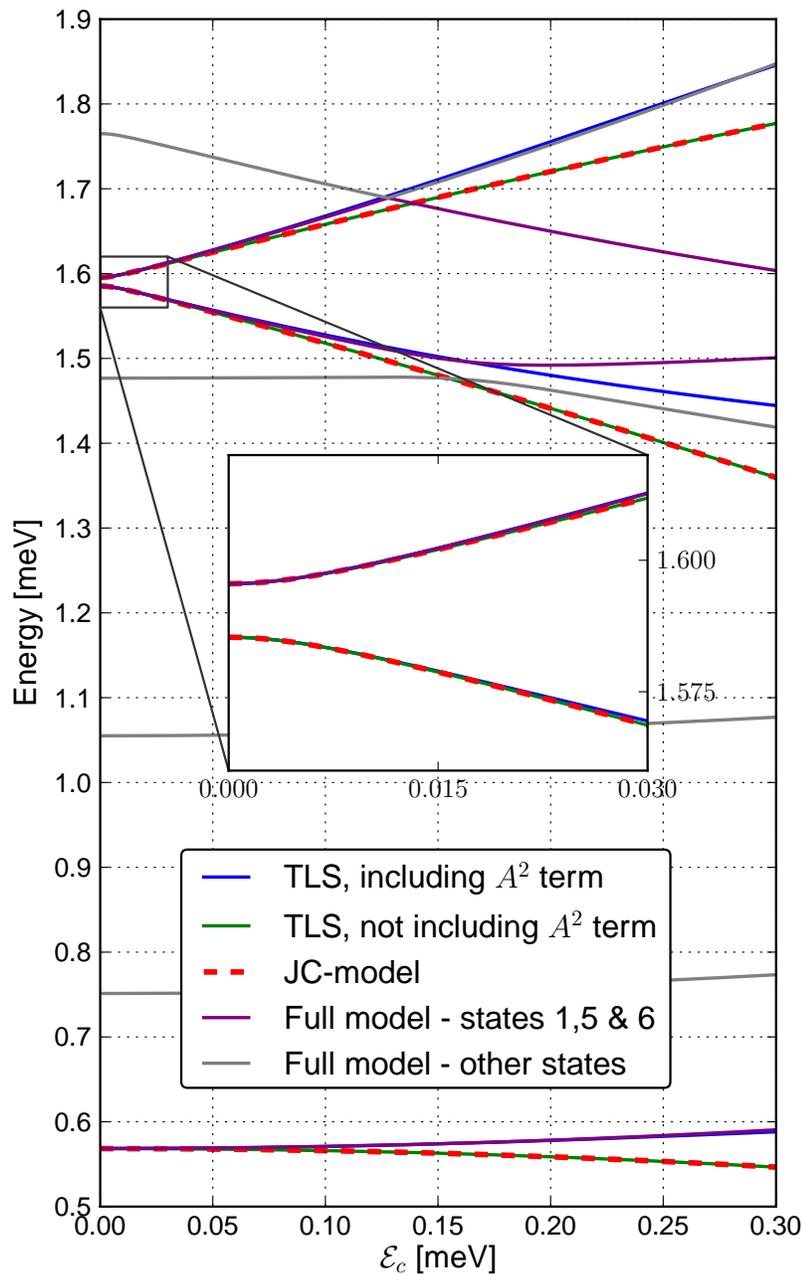}
      \caption{(Color online) Comparison of the many-body energy spectra
      versus the coupling strength $\mathcal{E}_\mathrm{c}$
      for the case of $\textrm{TE}_{101}$ mode ($y$-polarization).
      These energy states are obtained by TLS model including the $A^2$ term (blue),
      not including the $A^2$ term (green) and the JC approximation without magnetic field
      (red-dashed). The TLS model results are compared with the full numerical calculation for the
      compared lowest active levels $1$, $5$, and $6$ (purple) as well as inactive levels (gray).
      Other parameters are $B=0.1$ T, $\hbar\Omega_0=1.0$~meV, 
      $|g_{1,5}|=0.290$, and $\hbar\omega = 1.027$~meV. The inset shows the validity of the JC model
      in the weak coupling limit.
               }
      \label{JC-ypol}
\end{figure}

In Fig.\ \ref{JC-ypol}, we compare the many-body energy spectra as a
function of the electron-photon coupling strength
$\mathcal{E}_\mathrm{c}$ when the electronic system is embedded in a
$\textrm{TE}_{101}$ mode ($y$-polarization).  Attributed to selection rules of the
transverse parabolic confinement, we select the active states
$|1\rangle$ and $|5\rangle$ to compare with the TLS approximation.
In the zero coupling limit $\mathcal{E}_\mathrm{c} = 0$, we consider
the ground state energy $E_\mathrm{1}^0 \simeq 0.568$~meV and the
excited state energy $E_\mathrm{5}^0 \simeq 1.585$~meV so that
the energy level spacing $\Delta E_{15} = 1.017$~meV. In addition, we assume the detuning
$\delta_{15} = 0.01\times \Delta E_{15}$ such that the
single-photon energy $\hbar\omega = \Delta E_{15} + \delta_{15}
= 1.027$~meV. Moreover, we see that the state associated with
a ground state electron absorbing one photon is around $E_1^1 =
E_\mathrm{1}^0 + \hbar\omega = 1.596$~meV.

Figure \ref{JC-ypol} displays energy spectra calculated using the different models. 
As with the $x$-polarization, the energy spectrum obtained by the JC model almost coincides with the TLS result without the $A^2$ 
term. The difference (not visible in Fig.\ \ref{JC-ypol}) is due to effects of the external magnetic field.
When the $A^2$ term is included, the energy spectrum is blue-shifted in the strong 
coupling regime. When the higher MBSs are involved in the electron-photon 
coupling (full model) there is good
agreement with the TLS including the $A^2$ term until inactive states 
(not included in the two-level approximation) start to have influence, such as
the energy crossing at $\mathcal E_c\simeq 0.13$~meV and anti-crossing at 
$\mathcal E_c\simeq 0.17$~meV shown in Fig.\ \ref{JC-ypol}.
In the weak coupling regime $\mathcal{E}_\mathrm{c} < 0.1\hbar\omega \simeq 0.1$~meV, 
the JC model is approximately valid. The ultrastrong-coupling regime 
$\mathcal{E}_\mathrm{c} > \hbar\omega \simeq 1.0$~meV is not shown in this figure.

\begin{figure}[htbq]
      \centering
      \includegraphics[width=0.50\textwidth,angle=0]{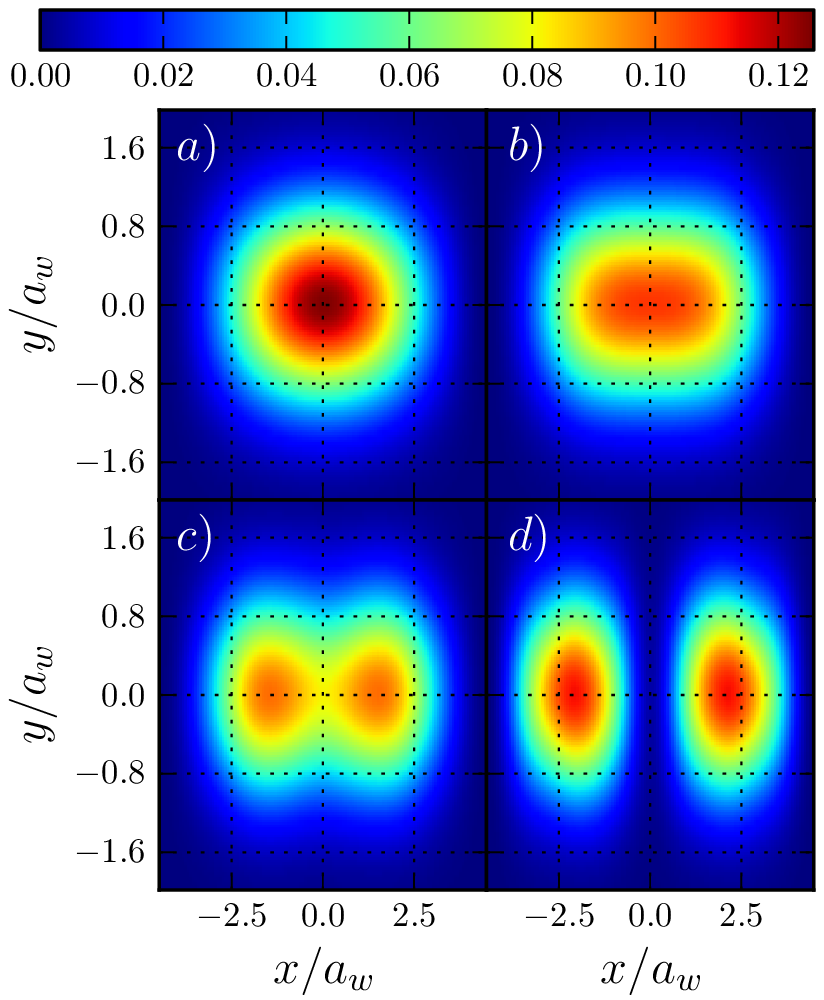}
      \caption{(Color online) Charge distribution (in units of $-e$) of the 
      third many-body state (see text for definition) in the case of
      $x$-polarization with electron-photon coupling strength
      $\mathcal{E}_\mathrm{c} =$ (a) $0.0$, (b) $0.2$, (c) $0.3$, and (d)
      $0.4$~meV.  The other parameters are the same with Fig.\ \ref{E-x}:
      $B=0.1$~T, $\hbar\Omega_0=1.0$~meV, and $\hbar\omega = 0.4$~meV.}
      \label{eQQ}
\end{figure}

The effects of the photon field on the charge distribution is illustrated in Fig.\ \ref{eQQ}, 
where the charge distribution $\langle Q({\mathbf r})\rangle$ of the third  
MBS is plotted for $x$-polarization for $\mathcal{E}_\mathrm{c}=0,0.2,0.3$ and $0.4$~meV. 
There is an energy crossing between the third and fourth state at $\mathcal E_c\simeq 0.16$ 
(see Fig.\ \ref{E-x}) so it is important to note that for $\mathcal E_c>0.16$, 
the third state refers to the fourth state counting from the bottom in Fig.\ \ref{E-x}. 
In Fig.\ \ref{eQQ}(a), the energy of the MBS is $E_0^1 \simeq 0.96$~meV. 
There is no coupling between the photons and electrons so the charge density 
is identical to that of the ground state labeled by the energy $E_0^0$. 
For $\mathcal{E}_\mathrm{c} =0.2$~meV$= \hbar\omega/2$, the charge distribution 
is stretched in the $x$-direction. This trend continues with increasing 
coupling strength. For $\mathcal E_{\rm c}=0.3$~meV the charge distribution 
starts to separate into two peaks and at $\mathcal{E}_\mathrm{c}=0.4$~meV  
the two peaks are completely separated. In other words, in the ultrastrong coupling regime  
$\mathcal{E}_\mathrm{c} = \hbar\omega$, a clear dipole-like charge distribution profile 
is observed. For the $y$-polarization, not shown here, the polarization of the
charge is much smaller due to the large value of the confinement energy $\hbar\Omega_0=1.0$ meV
compared to the photon energy $\hbar\omega =0.4$ meV. The system is anisotropic at 
the energy scale we employ here.

To summarize this section, we remind the reader that the JC
model exhibits energy spectrum with some levels decreasing when the
electron-photon coupling is increased~\cite{Feranchuk1996,Li2009}.
The energies may be negative if the electron-photon coupling is
very strong~\cite{Feranchuk1996}.  Comparing with the full numerical
calculation, it is thus unambiguous that the diamagnetic $A^2$
contribution, as well as higher energy electron states have to be included in 
the ultrastrong-coupling regime. We have done a simple literature search for
``ultrastrong'' and ``circuit-QED''. One has to keep in mind that 
authors have not always used the same definition for strong or ultrastrong,
but none found in our search results included the $A^2$ term in their 
model nor went beyond the two-level approximation.

\section{Concluding Remarks}\label{Sec:IV}

We have performed a numerical calculation of a microscopic model
describing a hybrid structure consisting of an electronic
nanostructure embedded in a cavity resonator. We have demonstrated
strong coupling features of Coulomb interacting electrons and photons
in a nanostructure embedded in a cavity resonator in an external
magnetic field.  The two-dimensional electronic nanostructure is
parabolically confined in the $y$-direction and hard-wall
confined in the $x$-direction that is embedded in a rectangular
photon cavity with a TE-mode electromagnetic field that may be
either $x$- or $y$-polarized. We have found that the
many-body energy spectrum is more sensitive to the photon field with
$x$-polarized electric component than that with
$y$-polarization for the selected geometry.  The system is
anisotropic in the energy range explored.

We have established that the diamagnetic $A^2$ term in the
Hamiltonian may provide a blue-shift correction to the energy
spectrum. However, when higher many-body states are included beyond
a two-level approximation, the results of the full numerical
calculation exhibit a smaller red-shift correction. This implies
that the two lowest levels become more stable when the higher energy
levels are included in the electron-photon coupled system. When the
$A^2$ term is not included the energy spectrum is lowered when
the coupling strength is increased, but the opposite trend is found
when the $A^2$ term is included in the calculation.

The widely employed two-level system approximation has been
reexamined comparing to results of our full numerical calculation
model. Qualitative difference of the energy spectrum between the JC
model and the full numerical calculation is found in the
strong-coupling regime. The JC-model includes no information about
the charge distribution of the system. A strong cavity photon field
can cause large polarization of the charge distribution, an effect 
seen in Fig.\ \ref{eQQ}. The reason for the high number of 
single-electron states needed in our full calculation is exactly this
large polarizing effect of the photon field. QED modeling of a circuit element
on the nanoscale in the ultrastrong coupling regime requires approximations 
beyond the JC-model or more general two level models.

In summary, we have presented a model adequate for accurate
numerical calculation for the electron-photon coupled energy
spectrum that is essential and will be utilized to explore 
time-dependent transport of electrons through a photon cavity
in a forthcoming publication~\cite{EM-GME}.


%
%
\ack
      This work was supported by the Icelandic Research and Instruments Funds,
      the Research Fund of the University of Iceland. CST is grateful to
      support from the National Science Council in Taiwan under Grants
      No.\ NSC97-2112-M-239-003-MY3 and No. NSC100-2112-M-239-001-MY3. HSG
      acknowledges support from the National Science Council, Taiwan,
      under Grants No.\ 97-2112-M- 002-012-MY3, and No.\
      100-2112-M-002-003-MY3, support from the Frontier and Innovative
      Research Program of the National Taiwan University under Grants No.
      99R80869 and No.\ 99R80871, and support from the focus group program
      of the National Center for Theoretical Sciences, Taiwan.
%
%

%
%
\section*{References}
\bibliographystyle{unsrt}

%
%
%
\end{document}